\documentstyle[11pt,epsfig]{article}

\begin{document}
\begin{titlepage}
\begin{flushright}
DFTT 30/98\\
DFCAL-TH 4/98\\
June 1998
\end{flushright}
\vskip0.5cm
\begin{center}
{\Large\bf 
Critical behavior of $3D$ $SU(2)$ gauge theory at finite temperature: exact
results from universality
}\\ 
\end{center}
\vskip 0.6cm
\centerline{R. Fiore$^a$, F. Gliozzi$^b$ and P. Provero$^b$}
\vskip 0.6cm
\centerline{\sl $^a$ Dipartimento di Fisica, Universit\`a della Calabria}
\centerline{\sl Istituto Nazionale di Fisica Nucleare, Gruppo collegato
di Cosenza}
\centerline{\sl Rende, I--87030 Cosenza, Italy
\footnote{e--mail: fiore@fis.unical.it}}
\vskip0.2cm
\centerline{\sl $^b$ Dipartimento di Fisica
Teorica dell'Universit\`a di Torino}
\centerline{\sl Istituto Nazionale di Fisica Nucleare, Sezione di Torino}
\centerline{\sl via P.Giuria 1, I--10125 Torino, Italy
\footnote{e--mail:gliozzi, provero@to.infn.it}}
\vskip 0.6cm
\begin{abstract}
We show that universality arguments, namely the Svetitsky--Yaffe conjecture,
allow one to obtain exact results on the critical behavior of 
$3D$ $SU(2)$ gauge theory at the finite temperature deconfinement transition,
through a mapping
into the $2D$ Ising model. In particular, we consider the 
finite--size scaling behavior of the plaquette
operator, which can be mapped into the energy operator of the $2D$ Ising model.
We obtain exact predictions for the dependence of the plaquette expectation
value on the size and shape of the lattice and we compare them to Monte Carlo
results, finding complete agreement. We discuss the application of this method
to the computation of more general correlators of the plaquette operator at
criticality, and its relevance to the study of the color flux tube structure.
\end{abstract}
\vskip .5cm
\hrule
\vskip.3cm
\noindent
$^{\ast}${\it Work supported by the Ministero italiano dell'Universit\`a 
e della Ricerca Scientifica e Tecnologica}
\end{titlepage}
\setcounter{footnote}{0}
\def\thefootnote{\arabic{footnote}}
\section{Introduction}
The idea of universality plays a major role in the modern understanding of
critical phenomena. All the physical systems undergoing a continuous phase
transition are believed to fall in a certain number of universality classes,
depending on the dimensionality of the system and its symmetry group, but not on
the details of the microscopic interactions (as long as these are
short--ranged). All the systems in a given universality class display the same
critical behavior, meaning that certain dimensionless quantities have the same
value for all the physical systems in a given class. Critical indices and
universal amplitude ratios are examples of these universal quantities.
\par
For gauge theories with a high temperature deconfinement transition, the
universality hypothesis takes the form of the Svetitsky--Yaffe conjecture,
\cite{sv-ya}, which can be formulated as follows: suppose
a $d+1$--dimensional gauge theory with gauge group $G$ has a second--order
deconfinement transition at a certain temperature $T_c$; consider the
$d$--dimensional statistical model with global symmetry group coinciding with
the center of the gauge group: if also this model displays a second--order phase
transition, then the two models belong to the same universality class. The
validity of the conjecture has been well established in several
Monte Carlo analyses (see {\em e.g.} \cite{ca-ha,teper} and references therein).
For the case we are interested in here, namely $G=SU(2)$ and $d=2$, 
a precise numerical test of the Svetitsky--Yaffe
conjecture can be found in Ref.\cite{teper}.
\par
The most obvious application of universality arguments, and in
particular of the
Svetitsky--Yaffe conjecture, is the prediction of the critical indices. 
For example, consider $SU(2)$ gauge theory; it undergoes a high--temperature
deconfinement transition which is known to be second--order in both three and
four space--time dimensions. The center of $SU(2)$ is $Z_2$, therefore the
dimensionally reduced statistical model is the Ising model, which 
has a second--order phase transition both in $d=2$ and $d=3$. Hence the
Svetitsky--Yaffe conjecture applies, and we can predict the critical indices 
of the $SU(2)$ deconfinement transition in $d+1$ dimensions to coincide with the
ones of the Ising model in $d$ dimensions.
\par
However, the predictive power of universality 
is certainly not limited to the values of the
critical indices. In Ref.\cite{gl-vi,gl,gl-pr} a program has been initiated of 
systematic exploitation of universality arguments in studying the
non--perturbative physics of gauge theories. For example, it was shown that
non--trivial results on finite--size effects and correlation functions 
at the deconfinement point can be obtained from
universality arguments. In this way it has been possible to evaluate exactly
the expectation value of the plaquette operator in presence of static sources, 
giving some new insight into the structure of the color flux tube for mesons 
and baryons.
\par
In this paper we continue the program 
by analysing the finite--size scaling behavior of the plaquette operator in $3D$
$SU(2)$ gauge theory at the deconfinement temperature. Since the $2D$ 
Ising model is exactly solved,
the Svetitsky--Yaffe conjecture gives in this case exact predictions on
finite--size scaling effects. We write down these predictions for the
expectation value of the plaquette operator and we compare them with Monte Carlo
results.
The same analysis was performed
in Ref.\cite{gl-pr} for $Z_2$ gauge theory.
\section{Finite--size behavior of the plaquette expectation value}
The Svetitsky--Yaffe conjecture can be seen as a mapping between observables of
the $3D$ $SU(2)$ gauge theory at finite temperature 
and operators of the $2D$ Ising model. The
Polyakov loop is mapped into the magnetization, while the plaquette operator
is mapped into a linear combination of the identity and the energy operators of
the statistical model \cite{gl-pr}:
\begin{equation}
\langle \Box \rangle =c_{1} \langle 1\rangle
+c_{\epsilon}\langle\epsilon\rangle+\dots
\end{equation}
where the expectation value in the l.h.s. is taken in the gauge theory, while 
the ones in the r.h.s. refer to the two--dimensional Ising model. The dots
represent contributions from secondary fields in the conformal families of the
identity and energy operators, whose contributions are subleading for
asymptotically large lattices.
\par
The finite--size dependence of the energy expectation value in the
two--dimensional Ising model on a torus is \cite{fe-fi,di-sa-zu,it-dr}
\begin{equation}
\langle \epsilon\rangle = 
\frac{\pi\sqrt{\Im m \tau}\left|\eta(\tau)\right|^{2}}
{\sqrt{A}Z_{1/2}(\tau)}\label{fse}
\end{equation}
where $A\equiv L_1 L_2$ and $\tau\equiv iL_1/L_2$ 
are respectively the area and the modular 
parameter of the torus. $Z_{1/2}$ is the partition function of the Ising model 
at the critical point:
\begin{equation}
Z_{1/2}=\frac{1}{2}\sum_{\nu=2}^{4}\left|\frac{\theta_{\nu}(0,\tau)}
{\eta(\tau)}\right|\ \ .
\end{equation}
(We follow the notations of Ref.\cite{it-dr} for the Jacobi theta functions
$\theta_{\nu}$ and the Dedekind function $\eta$).
\par
Consider now $3D$ $SU(2)$ lattice gauge theory regularized on a $L_1\times
L_2\times N_t$ lattice, with $L_1\ ,L_2\gg N_t$. 
For a given $N_t$ the gauge coupling $\beta$ can be
tuned to a critical value $\beta_c(N_t)$ to simulate the theory at the
finite temperature deconfinement phase transition. Precise evaluations of
$\beta_c(N_t)$ for various values of $N_t$ are available in the literature 
\cite{teper}. The universality argument gives us the 
following prediction for the finite--size scaling
behavior of the plaquette operator at the deconfinement point
\begin{equation}
\langle \Box \rangle_{L_{1}L_{2}}=c_{1}+c_{\epsilon}\frac{F(\tau)}
{\sqrt{L_{1}L_{2}}}+O(1/L_1 L_2)\label{fseplaq}
\end{equation}
where $F$ is a function of the modular parameter $\tau\equiv iL_1/L_2$
only:
\begin{equation}
F(\tau)=\frac{\pi \sqrt{\Im m\tau}\left|\eta(\tau)\right|^2}
{Z_{1/2}(\tau)}\ .
\end{equation}
Here $c_1$ and $c_\epsilon$ are non--universal constants which depend on $N_t$
and must be determined numerically. 
Once these have been determined, Eq.~(\ref{fseplaq}) predicts the
expectation value of the plaquette operator for all sizes and shapes of the
lattice, {\em i.e.} for all values of $L_1$ and $L_2$. 
The $O(1/L_1L_2)$ corrections represent the contribution of secondary fields.
Therefore Eq.~(\ref{fseplaq}) is valid asymptotically for large lattices.
\section{Plaquette correlators}
Once the constants $c_1$ and $c_\epsilon$ have been determined at a given value
of $N_t$, for example through the finite--size scaling analysis presented here,
all the correlation functions of the plaquette operator at criticality
are in principle exactly
computable, since they can be readily derived from the corresponding correlators
of the energy operator in the $2d$ Ising model. 
\par
Consider for example the plaquette expectation value in presence of static
sources, {\em i.e.} the correlation function of the plaquette with $n$ Polyakov
loops:
\begin{equation}
G(x;y_1,\dots,y_n)=\langle\Box(x)P(y_1)\dots P(y_n)\rangle
\end{equation}
This is the typical quantity to study if one is interested in the structure of
the color flux tube, since it represents the density of action in presence of
external sources. 
Universality then tells us that the connected part of $G(x;y_1,\dots,y_n)$
is given by
\begin{equation}
G_c(x;y_1,\dots,y_n)=c_\sigma^n c_\epsilon
\langle \epsilon(x)\sigma(y_1)\dots\sigma(y_n)\rangle
\end{equation}
where the correlator on the r.h.s. is computed in the critical $2D$ Ising model,
$\sigma$ is the spin operator and $c_\sigma$ is the constant which relates it to
the Polyakov loop; $c_\sigma$ can be determined with methods similar to the ones
we used here to determine $c_\epsilon$.
\par
Therefore the Svetitsky--Yaffe conjecture, together with the exact results of
$2D$ conformal field theory, gives us complete control over the critical
behavior of $3D$ $SU(2)$ gauge theory at the finite temperature
deconfinement transition. More generally, this holds for every $3D$ gauge theory
whose deconfinement transition is second--order, including $SU(3)$. Some studies
of the flux--tube structure following this line were presented in Ref.
\cite{gl-pr}.
\section{Comparison with Monte Carlo results for $SU(2)$ gauge theory}
Three--dimensional finite temperature gauge theories are simulated by using 
$L_1\times L_2 \times N_t$ lattices with $N_t\ll L_1,\ L_2$. We have performed
our simulations of $SU(2)$ pure gauge theory with $N_t=2,4$ and $L_1, L_2$ 
varying between $8$ and $30$, and  modular parameter $\Im m \tau$ 
between 1 and 3. 
The boundary conditions are periodic in all three directions.
For each value of $N_t$ we have chosen the
coupling $\beta_c(N_t)$ corresponding to the deconfinement transition point:
from Ref.\cite{teper} we have
\begin{eqnarray}
\beta_c(2)&=&3.469\\
\beta_c(4)&=&6.588 
\end{eqnarray}
The only free parameters in Eq.~(\ref{fseplaq}) are the non--universal
constants $c_1$ and $c_\epsilon$, which depend on the value on $N_t$ and on the
kind of plaquette we are considering (time--like or space--like).
Therefore we have a total of four sets of data to be compared with 
the theoretical prediction: 
for each of these sets we can perform a two--parameter fit
with Eq.~(\ref{fseplaq}). 
It turns out however that finite--size effects on space--like plaquettes are
of the same order of magnitude 
than the typical statistical uncertainties of our simulations. 
Therefore from now on we will consider time--like plaquettes only.
\par
It is important to stress that for each value of $N_t$ all the data,
corresponding to different values of $\Im m\tau$, are included in the same 
two--parameter fit. Specifically, for each lattice with sides $(L_1,L_2)$
we define an "effective area"
\begin{equation}
\alpha(L_1,L_2)=\frac{L_1 L_2}{F^2\left(i\frac{L_1}{L_2}\right)}
\end{equation}
so that Eq.~(\ref{fseplaq}) becomes
\begin{equation}
\langle \Box \rangle_{L_{1}L_{2}}=c_{1}+\frac{c_{\epsilon}}{\sqrt{\alpha(L_1,
L_2)}}\label{fsea}
\end{equation}
and we can fit the plaquette expectation value to a linear function of
$1/\sqrt{\alpha}$.
\par
The Monte Carlo results for the plaquette expectation values are reported in
Tab.1 and Tab.2 while
the results of the fits for $N_t=2$ and $N_t=4$ are reported in Tab.3. 
The values
of the reduced $\chi^2$ show that the agreement is very satisfactory. 
In Figs. 1 and 2 the plaquette expectation values are plotted against 
$1/\sqrt{\alpha}$ together with the best fit line. The same data are plotted
against $1/\sqrt{L_1L_2}$ in Fig. 3 for $N_t=2$. 
This shows the crucial importance 
of including the non--trivial part of Eq.~(\ref{fseplaq}), namely its
$\tau$--dependence.

\par 
The same analysis was performed in Ref.\cite{gl-pr} for $3D$ $Z_2$ gauge
theory, whose finite temperature deconfinement transition is also in the
universality class of the $2D$ Ising model. 
\footnote{The values of $c_\epsilon$ reported in Ref.\cite{gl-pr} for $Z_2$
gauge theory are negative: this is due to the fact that the simulation was
actually performed in the $3D$ spin Ising model, using  $3D$ duality.
As a consequence, the Svetitsky-Yaffe mapping between $3D$ observables  and 
$2D$ Ising operators at criticality includes a duality transformation, which 
changes the sign of $\epsilon$.}
\section{Conclusions}
We have shown that the Svetitsky--Yaffe conjecture, {\em i.e.} universality
applied to the deconfinement transition, provides an exact description of
finite--size effects for $3D$ $SU(2)$ gauge theory at the deconfinement
point. This description is obtained by mapping the gauge theory into the $2D$
Ising model, which is in the same universality class, and is exactly solved.
We have compared the predictions obtained from this mapping with Monte Carlo
results, finding complete agreement.
\par
The main motivation for this work was not to verify the validity of the
Svetitsky--Yaffe conjecture, which is by now well established. 
Our intent is rather to stress that universality arguments 
provide a powerful,  analytical approach to a deeply non--perturbative region, 
namely the
deconfinement transition and its neighborhood. This is especially true for
$3D$ gauge theories, since critical behavior in $2D$ is completely understood
with the techniques  of conformal field theory.
\vskip 1.5cm
\underline {Acknowledgement}: We thank P. Cea and L. Cosmai for providing 
us with their efficient Monte Carlo code. 
One of us (R.F) thanks also A. Papa for helpful comments.

\begin{table}[ht]
\caption{\sl
Monte Carlo results for the time--like 
plaquette expectation values at the deconfinement
transition point, for $N_t=2$
}
\label{mcrplaq2}
\begin{center}
\begin{tabular}{|c|c|c|}
\hline
$L_1$ &$L_2$ & $\langle \Box \rangle$ \\
\hline
10&10&0.690737(26)\\ \hline
10&20&0.689930(26)\\ \hline
10&30&0.689402(26)\\ \hline
12&12&0.690393(26)\\ \hline
12&24&0.689762(23)\\ \hline
12&28&0.689593(26)\\ \hline
12&36&0.689304(26)\\ \hline
14&14&0.690201(22)\\ \hline
14&28&0.689626(26)\\ \hline
16&16&0.690090(23)\\ \hline
16&20&0.689912(21)\\ \hline
16&32&0.689566(23)\\ \hline
18&18&0.689963(21)\\ \hline
20&20&0.689835(20)\\ \hline
22&22&0.689753(30)\\ \hline
24&24&0.689697(20)\\ \hline
\end{tabular}
\end{center}
\end{table}
\begin{table}[ht]
\caption{\sl
Same as Tab. 1 for $N_t=4$
}
\label{mcrplaq4}
\begin{center}
\begin{tabular}{|c|c|c|}
\hline
$L_1$ &$L_2$ & $\langle \Box \rangle$ \\
\hline
12&12&0.8418794(72)\\ \hline
14&14&0.8418684(64)\\ \hline
14&20&0.8418340(52)\\ \hline
16&16&0.8418488(57)\\ \hline
16&24&0.8418210(54)\\ \hline
20&20&0.8418219(49)\\ \hline
20&24&0.8418157(54)\\ \hline
22&22&0.8418143(43)\\ \hline
\end{tabular}
\end{center}
\end{table}
\begin{table}[ht]
\caption{\sl  
Results of the fit of the plaquette expectation values with Eq.~(\ref{fseplaq})
for $N_t=2$ and $N_t=4$. Each fit includes data from different values of
the modular parameter.
}
 \label{fit}
  \begin{center}
   \begin{tabular}{|c|c|c|c|}
   \hline
 $N_t$ & $c_1$ & $c_\epsilon$ & $\chi^2_{red}$ \\
   \hline
2 & 0.688923(18)&0.01856(33)&0.92\\
4 & 0.8417328(93)& 0.00186(17)& 0.32\\
   \hline
   \end{tabular}
  \end{center}
 \end{table}
\begin{figure}[ht]
\begin{center}
\mbox{\epsfig{file=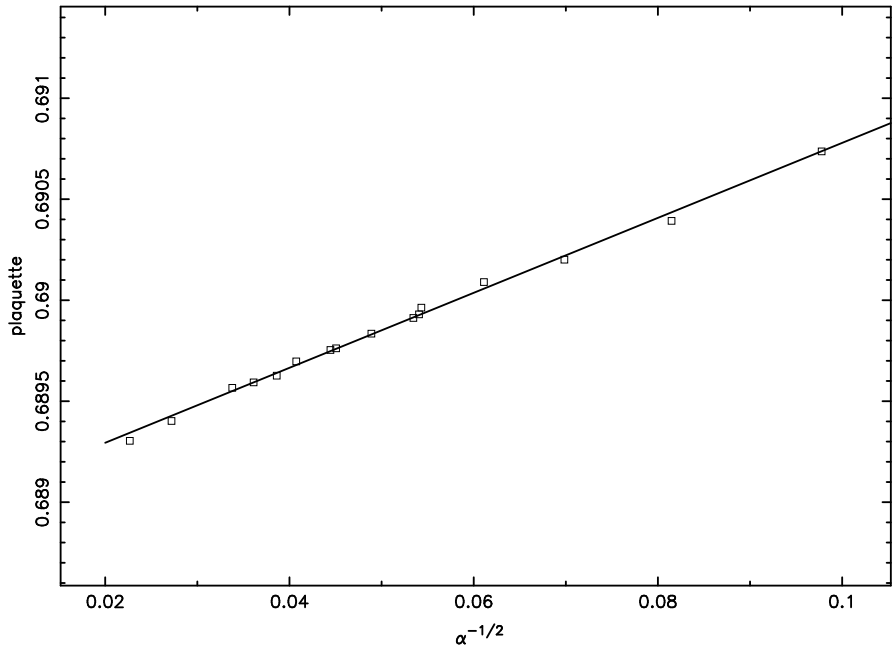}}
\vskip2.mm
\caption{Plaquette expectation value as a function of $\alpha^{-1/2}$
for $N_t=2$. Error bars are comparable to the size of the plotting
symbol.
The line is the best fit to Eq.~(\ref{fsea})}
\label{fig:nt2}
\end{center}
\end{figure}
\begin{figure}[ht]
\begin{center}
\mbox{\epsfig{file=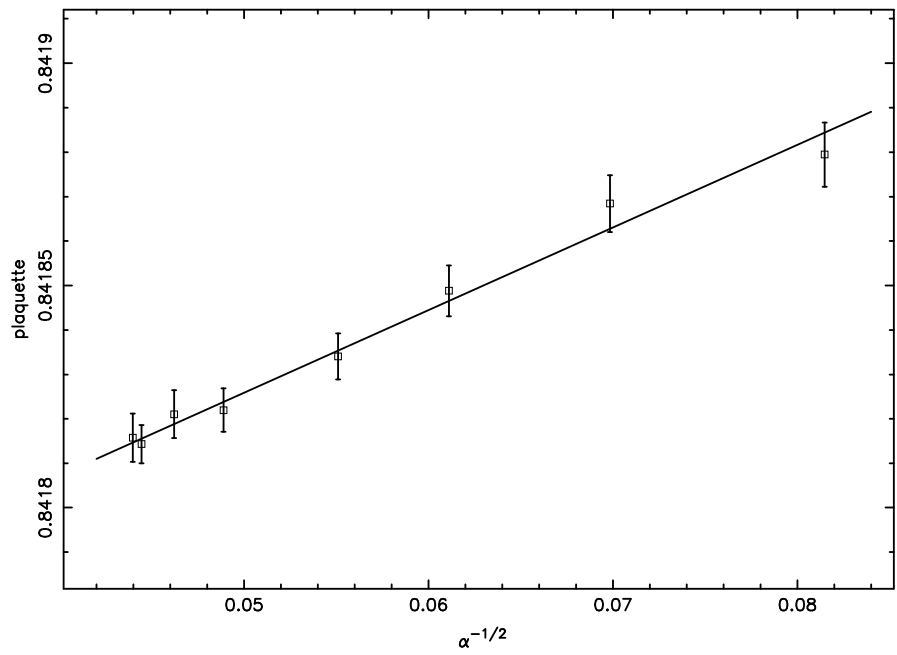}}
\vskip2.mm
\caption{
Same as Fig. 1 for $N_t=4$
}
\label{fig:nt4}
\end{center}
\end{figure}
\begin{figure}[ht]
\begin{center}
\mbox{\epsfig{file=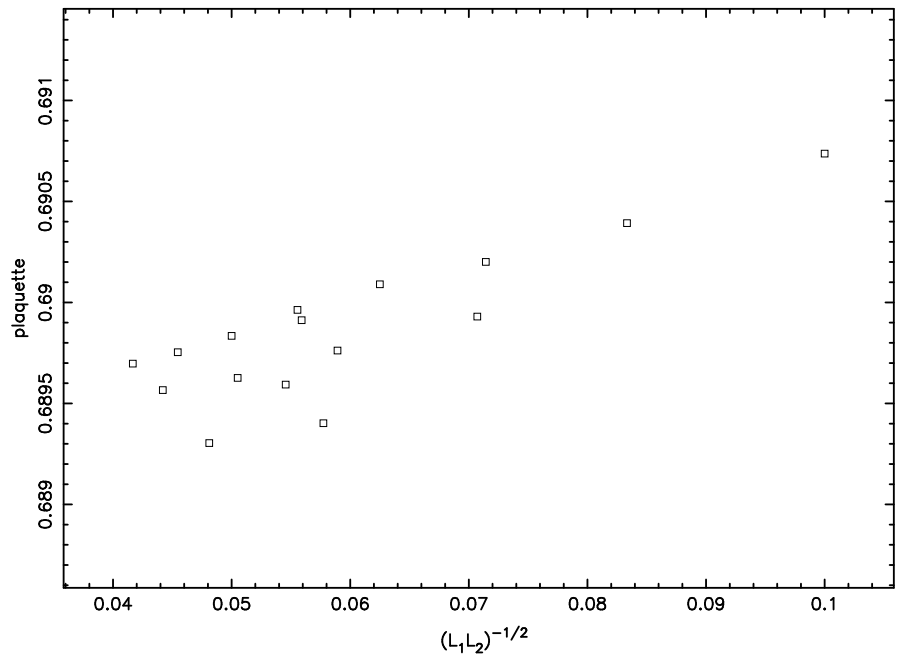}}
\vskip2.mm
\caption
{Plaquette expectation value as a function of $(L_1 L_2)^{-1/2}$
for $N_t=2$}
\label{fig:nt2w}
\end{center}
\end{figure}
\end{document}